\PassOptionsToPackage{unicode}{hyperref}
\PassOptionsToPackage{hyphens}{url}
\PassOptionsToPackage{dvipsnames,svgnames,x11names}{xcolor}
\documentclass[
  12pt]{article}

\usepackage{amsmath,amssymb}
\usepackage{iftex}
\ifPDFTeX
  \usepackage[T1]{fontenc}
  \usepackage[utf8]{inputenc}
  \usepackage{textcomp} 
\else 
  \usepackage{unicode-math}
  \defaultfontfeatures{Scale=MatchLowercase}
  \defaultfontfeatures[\rmfamily]{Ligatures=TeX,Scale=1}
\fi
\usepackage{lmodern}
\ifPDFTeX\else  
\fi
\IfFileExists{upquote.sty}{\usepackage{upquote}}{}
\IfFileExists{microtype.sty}{
  \usepackage[]{microtype}
  \UseMicrotypeSet[protrusion]{basicmath} 
}{}
\makeatletter
\@ifundefined{KOMAClassName}{
  \IfFileExists{parskip.sty}{%
    \usepackage{parskip}
  }{
    \setlength{\parindent}{0pt}
    \setlength{\parskip}{6pt plus 2pt minus 1pt}}
}{
  \KOMAoptions{parskip=half}}
\makeatother
\usepackage{xcolor}
\makeatletter
\ifx\paragraph\undefined\else
  \let\oldparagraph\paragraph
  \renewcommand{\paragraph}{
    \@ifstar
      \xxxParagraphStar
      \xxxParagraphNoStar
  }
  \newcommand{\xxxParagraphStar}[1]{\oldparagraph*{#1}\mbox{}}
  \newcommand{\xxxParagraphNoStar}[1]{\oldparagraph{#1}\mbox{}}
\fi
\ifx\subparagraph\undefined\else
  \let\oldsubparagraph\subparagraph
  \renewcommand{\subparagraph}{
    \@ifstar
      \xxxSubParagraphStar
      \xxxSubParagraphNoStar
  }
  \newcommand{\xxxSubParagraphStar}[1]{\oldsubparagraph*{#1}\mbox{}}
  \newcommand{\xxxSubParagraphNoStar}[1]{\oldsubparagraph{#1}\mbox{}}
\fi
\makeatother

\usepackage{longtable,booktabs,array}
\usepackage{calc} 
\usepackage{etoolbox}
\makeatletter
\patchcmd\longtable{\par}{\if@noskipsec\mbox{}\fi\par}{}{}
\makeatother
\IfFileExists{footnotehyper.sty}{\usepackage{footnotehyper}}{\usepackage{footnote}}
\makesavenoteenv{longtable}
\usepackage{graphicx}
\makeatletter
\def\maxwidth{\ifdim\Gin@nat@width>\linewidth\linewidth\else\Gin@nat@width\fi}
\def\maxheight{\ifdim\Gin@nat@height>\textheight\textheight\else\Gin@nat@height\fi}
\makeatother
\setkeys{Gin}{width=\maxwidth,height=\maxheight,keepaspectratio}
\makeatletter
\def\fps@figure{htbp}
\makeatother

\makeatletter
\@ifpackageloaded{caption}{}{\usepackage{caption}}
\AtBeginDocument{%
\ifdefined\contentsname
  \renewcommand*\contentsname{Table of contents}
\else
  \newcommand\contentsname{Table of contents}
\fi
\ifdefined\listfigurename
  \renewcommand*\listfigurename{List of Figures}
\else
  \newcommand\listfigurename{List of Figures}
\fi
\ifdefined\listtablename
  \renewcommand*\listtablename{List of Tables}
\else
  \newcommand\listtablename{List of Tables}
\fi
\ifdefined\figurename
  \renewcommand*\figurename{Figure}
\else
  \newcommand\figurename{Figure}
\fi
\ifdefined\tablename
  \renewcommand*\tablename{Table}
\else
  \newcommand\tablename{Table}
\fi
}
\@ifpackageloaded{float}{}{\usepackage{float}}
\floatstyle{ruled}
\@ifundefined{c@chapter}{\newfloat{codelisting}{h}{lop}}{\newfloat{codelisting}{h}{lop}[chapter]}
\floatname{codelisting}{Listing}

\makeatother
\makeatletter
\@ifpackageloaded{caption}{}{\usepackage{caption}}
\@ifpackageloaded{subcaption}{}{\usepackage{subcaption}}
\makeatother

\ifLuaTeX
  \usepackage{selnolig}  
\fi
\usepackage[]{natbib}
\usepackage{bookmark}

\IfFileExists{xurl.sty}{\usepackage{xurl}}{} 
\hypersetup{
  pdftitle={Title},
  pdfauthor={Author 1; Author 2},
  pdfkeywords={3 to 6 keywords, that do not appear in the title},
  colorlinks=true,
  linkcolor={blue},
  filecolor={Maroon},
  citecolor={Blue},
  urlcolor={Blue},
  pdfcreator={LaTeX via pandoc}}

\def\half{\hbox{$1\over2$}}

\def\la{(\lambda)}
\newcommand{\anon}{1}

\newtheorem{theorem}{Theorem}[section]
\newtheorem{lemma}[theorem]{Lemma}

\begin{document}

\def\spacingset#1{\renewcommand{\baselinestretch}%
{#1}\small\normalsize} \spacingset{1}


\if1\anon
{
\title{\bf On Stein's Method of Moments and Generalized Score Matching}
\author{Alfred Kume\\
Department of Statistics, University of Kent, UK\\
and\\
Stephen G. Walker\\
Department of Statistics, University of Kent, UK}
\maketitle
} \fi

\if0\anon
{
\bigskip
\bigskip
\bigskip
\begin{center}
{\LARGE\bf Title}
\end{center}
\medskip
} \fi

\bigskip
\begin{abstract}
The Stein class used in method of  moments parameter estimation has two functions which need to be specified, for which there is no persuasive arguments for any particular choice. We show that by setting one to be the derivative of the density score function with respect to the parameter leads to a generalized score matching estimator with a choice of weight function.  
However, choosing a suitable weight function for generalized score matching is not straightforward. 
We show the weight function is equivalent to a transform of the data and using a score estimator, with an optimal transform being to a normal sample, using for example Box-Cox. 
We compare our proposal with an alternative means by which to handle the weight function, which is to use a generalized method of moment estimator.
\end{abstract}

\noindent%
\vfill

\newpage
\spacingset{1.8} 

\section{Introduction}

The Method of Moments, \cite{quandt78}, is a strategy for the estimation of a parameter $\theta \in \mathbb{R}^p$ from the family of density functions $f(x\mid\theta)$ by solving
$n^{-1}\sum_{i=1}^n x_i^k-g_k(\theta)=0$,
for some integers $k$, where $g_k(\theta)=\int x^k\,f(x\mid\theta)\,dx$, and $x_{1:n}$ are a sample from the true density function with parameter value $\theta^*$. 
More generally, one uses a function $\lambda(x,\theta)$ satisfying
$\int \lambda(x,\theta)\,f(x\mid\theta)\,dx=0,$
for all $\theta$.
Here $\lambda$ can be a vector ($p$-dimensional) of functions. 

A particular class of moment function is provided by Stein's method of moments, see \cite{Stein72} and, for  example, \cite{Ebner}.
This class uses
\begin{equation}\label{eq:lambda}
\lambda(x,\theta)=(\tau_\theta(x)\,w(x)\,f(x\mid\theta))'\,\big/\,f(x\mid\theta), 
\end{equation}
for to be chosen functions $\tau_\theta(x)$ and $w(x)$. Here and throughout $'$ represents differentiation with respect to $x$. There is no clear guidance on the choice of these functions, though a recommendation (\cite{Ebner}) is that $\tau_\theta(x)$ is taken to be a Stein kernel of the form
$$\tau_\theta(x)=\int_x^\infty \{E(X)-y\}f(y\mid\theta)\,dy\,\big/\,f(x\mid\theta).$$
If $\theta$ is $p$-dimensional the usual procedure is to obtain $p$ equations by using $p$ functions $\underline{w}=(w_{j})_{j=1:p}$ so that (\ref{eq:lambda}) is represented by $p$ equations. The $\tau_\theta$ is taken to be a one-dimensional function. 

More generally, where any single choice of $\underline{w}$ could be seen as lacking motivation, one could select a number of different $\underline{w}$ functions and to somehow combine outputs from each into a single estimator. 
This is the idea behind generalized method of moments (GMM), see \cite{Hansen82}. Here we detail how to obtain the GMM estimator, see for example, \cite{Andrews}. In general, suppose we use functions $\lambda=(\lambda_j)_{j=1:M}$. From each function we obtain the usual estimator $\widehat\theta_j$ by setting
$\sum_{i=1}^n \lambda_j(x_i,\theta)=0.$
To obtain a weight matrix, we construct the matrix $W$ with
$(j,k)$ entry the sample covariance between $n$-vectors
$z_j$ and $z_k$, where 
$z_j=(\lambda_j(x_i),\theta)_{i=1:n}$.
The GMM estimator is obtained by minimizing
$G(\theta)=z(\theta)'\,W^{-1}\,z(\theta)$
where $z(\theta)=(z_1(\theta),\ldots,z_M(\theta))$
and
$z_j(\theta)=\sum_{i=1}^n \lambda_j(x_i,\theta).$
Properties of the estimator and related information can be found in \cite{Hall05}.

A different approach to estimation is a generalized score matching estimator, see \cite{Yu19}, \cite{Scealy23}, and \cite{Xu25}, which minimizes the empirical estimation of the weighted Fisher information distance given by
\begin{equation}\label{fwdist}
d_F(f,f_\theta)=\int f(x)\, w(x)\,\left(s(x)-s_\theta(x)\right)^2\,dx,
\end{equation}
where $f$ and $s$ denote the true density and score functions, respectively and $s_\theta$ is the usual score function $(\partial/\partial x)\log f(x\mid\theta)$. Taking $w\equiv 1$ yields the score matching, or Hyvarinen estimator, see \cite{Hyvarinen05}. Here the choice of the weight function $w$ is also problematic.

Our contribution is as follows and for convenience and clarity we list this as a sequence with the aim being to motivate a particular choice of $\tau_\theta$ and $w$ from (\ref{eq:lambda}): 

\begin{description}

\item 1. Using the Stein class (\ref{eq:lambda}) we obtain the necessary $p$ equations by taking $w$ to be a one dimensional function and take $\tau_\theta(x)$ to be 
$\tau_\theta(x)=\nabla s_\theta(x)=(\partial/\partial\theta) s_\theta(x)$ and so is a $p$-vector of functions.  

\item 2. We show that with this choice of $\tau_\theta$ the Stein method of moment estimator coincides with the generalized score matching estimator with weight function $w$.

\item 3. We show that the generalized score matching estimator is equivalent to a score matching  estimator using the transformation of the data
$Y=g(X)$ with $w(x)=(h'(g(x)))^2$ where $h=g^{-1}$.

\item 4. We argue that the optimal transformation is to produce the $Y$ sample as close to normal as possible. This can be achieved using the Box-Cox class, \cite{Box64}, for example. The heuristic for now is that the score matching estimator, which comes from a normal sample and model, is precisely the MLE.

\end{description}

\noindent
Putting these arguments together, and exploiting the new connection between the Stein class and generalized score matching, we propose the GMM can be replaced by the optimal transform and weight function $w$ and the use of the score matching estimator. Simulation studies show that the transformed data score matching estimator performs at least as well as, if not better, than the corresponding GMM estimator.

Note that recently, \cite{Barp19}, has shown that a score matching estimator is a special case of a Stein discrepancy estimator, see also \cite{Oates22}. Note that Stein discrepancy estimators are \underline{not} the same as Stein method of moment estimators. The former uses a discrepancy between distributions while the  latter solves moment equations.

In Section 2 we describe the score matching and the generalized score matching estimation approaches and show they are connected by a transformation of the data.
In Section 3 we elaborate on the connection between Stein's method of moment estimator and the generalized score matching approach. 
We also provide motivation for the Stein class method of moment approach using the $w$ function to follow from a transformation of the data to a normal looking sample, using, for example, the Box-Cox family of transformations.
Section 4 looks at the general set up and equations for the implementation of the GMM, our aim is to compare the GMM approach with the ``optimal" approach of using the $w$ function based on a normal transformation. Section 5 presents a simuation study.

\section{Score and generalized score matching}

Models for which a maximum likelihood estimator (MLE) is difficult to obtain, due to an intractable normalizing constant, can be estimated using an alternative strategy based on score functions. The MLE is based on minimizing a Kullback-Leibler divergence between the family of density functions $f(x\mid\theta)$ indexed by the parameter $\theta\in\Theta$ and the true density function. With the true density being unknown, a sample Kullback-Leibler divergence estimator using the empirical measure is used instead, and this leads to the MLE. Other distances can be used provided it is possible to substitute the empirical for the true density and obtain a Monte Carlo approximation to the distance or divergence.

For ease of notation, we write $f(x\mid\theta^*)$, the true density function with true parameter $\theta^*$, as $f(x)$, and $s_{\theta^*}(x)$ as $s(x)$.
The Fisher Information distance, see, for example, \cite{Atkinson1981}, given by
$$d_F(\theta^*,\theta)=\int f(x\mid\theta^*)\big(s_{\theta^*}(x)-s_\theta(x)\big)^2\,dx,$$
allows for parameter estimation via a  Monte Carlo approximation to the distance. Since it only depends on score functions,  any intractable normalizing constant conveniently disappears.

The idea is to find the $\theta$ minimizing
\begin{equation}\label{score}
D(\theta)=\int f(x)\,\left\{s(x)-s_\theta(x)\right\}^2\,dx,
\end{equation}
and this distance will be approximated using the sample $x_{1:n}$. To this end, write
$D(\theta)$ in the form in which it only depends on the $\theta$, i.e.
$$D(\theta)=\int f(x)\,s^2_\theta(x)\,dx-2\int f'(x)\,s_\theta(x)\,dx.$$
While the first term is set up for a Monte Carlo estimator, the second is not, and an integration by parts is required.
For this to work, 
$f(x)\,s_\theta(x)$
needs to vanish at the boundary points for all $\theta$.
Assuming this to be true, then,
$D(\theta)=\int f(x)\, s^2_{\theta}(x)\,dx+2\int f(x)\,s^{'}_{\theta}(x)\,dx,$
which is approximated from the sample as
\begin{equation}\label{eq:dhat}
\widehat{D}(\theta)=n^{-1}\,\sum_{i=1}^n \left\{s^2_\theta(x_i) +2\,s'_\theta(x_i) \right\}.
\end{equation}
The score estimator minimizes $\widehat{D}(\theta)$ and this approach to parameter estimation was proposed by \cite{Hyvarinen05}.

More recently, motivated by constrained space estimation problems, weight functions have been introduced, with  
(\ref{score}) being generalized to
\begin{equation}\label{wscore}
D(\theta)=\int f(x)\,w(x)\,\left\{s(x)-s_\theta(x)\right\}^2\,dx.
\end{equation}
See, for example, \cite{Yu19}, \cite{Scealy23} and \cite{Xu25}.

Here we show that (\ref{wscore}) can be shown to arise from (\ref{score}) by a transformation.
If $s_X(x)$ is the score function for variable $X$ and we transform $Y=g(X)$, then straightforward calculations give
$S_Y(y)=S_X(h(y))\,h'(y)+(\,\,\log |h'(y)|)',$
where $h\equiv g^{-1}$. Hence, (\ref{score}) becomes
(\ref{wscore}) with $w(x)=(h'(g(x)))^2$.
Note that this is invariant to linear transforms, i.e. if $\widetilde{g}=\alpha\,g+\beta$ then it is easy to show that $\widetilde{w}=w$.

By applying a similar argument for the boundary conditions, the minimizer of \eqref{wscore} is estimated by minimizing
\begin{equation}
\widehat{D}_w(\theta)=\sum_{i=1}^n \left\{s^2_{\theta}(x_i) w(x_i)+2(s_{\theta}(x_i) w(x_i))^{'}\right\}.
\label{wscore1}
\end{equation}
Within this framework, the choice of function $w(x)$ is problematic.  
On the other hand, we can link up the generalized score matching approach to a method of moment estimator and then use GMM to determine optimal weights based on a set of functions. 


\section{Method of moments and score functions}

In this section we provide a new perspective on the score matching approach and its connection with method of moments estimation; i.e. (\ref{eq:lambda}),
for some differentiable function $g_\theta$.  It turns out (see Lemma 3.2) that the condition for minimizing the Fisher distance (\ref{score}) and its weighted version (\ref{wscore}) can be expressed in terms of such functions.


\begin{lemma} (Stein Method of Moments). If
$g_\theta(x) \,f(x\mid\theta)$ vanishes at the boundary values then
$\int f(x\mid\theta)\,\lambda_{g}(x,\theta)\,dx=0,$
where $\lambda_{g}(x,\theta)=(g_\theta(x)\,f(x\mid\theta))'/f(x\mid\theta)$.
\end{lemma}

\noindent
{\sl Proof}.
Now $f(x\mid\theta)\,s_{\theta}(x)=f'(x\mid\theta)$ and so
the integral becomes
$$\int \{g_\theta(x)\,f'(x\mid\theta)+g_\theta'(x)\,f(x\mid\theta)\}\,dx =\int (g_\theta(x)f(x\mid\theta))'\,dx=0$$
which proves the lemma.
\hfill$\Box$

\vspace{0.1in}
\noindent
Note that because $\int \{g_\theta(x)\,f'(x\mid\theta)+g_\theta'(x)\,f(x\mid\theta)\}\,dx=0$, we have that
$
E\{g_\theta(X)s_\theta(x))=-E(g_\theta'(X)\}
$.
In the exponential family case, $s_\theta(x)$ is linear in $\theta$ which leads to explicit solutions.
We now make the connection, through suitable choices of $g_\theta(x)$, with score matching, by considering the first order derivative equation of \eqref{wscore1}. 

\begin{lemma}
If $g_\theta(x)=\nabla s_\theta(x)$ then the method of moment estimator coincides with the estimator from (\ref{eq:dhat}). If $g_\theta(x)=w(x) \nabla s_\theta(x)$ then the method of moment estimator coincides with the estimator from (\ref{wscore1}).
\end{lemma}

\noindent
{\sl Proof}.
The generalized score matching estimator minimizes
$$\sum_{i=1}^n \left\{w(x_i)\, s^2_\theta(x_i)+2(w(x_i)\,s_\theta(x_i))'\right\}.$$
Differentiating with respect to $\theta$, we set to zero
$$\sum_{i=1}^n\left\{ w(x_i)\,s_\theta(x_i)\,\nabla s_\theta(x_i)+(w(x_i)\,\nabla s_\theta(x_i))'\right \}=0.$$
This is equivalent to $\sum_{i=1}^n \lambda(x_i,\theta)=0$
where $\lambda(x,\theta)$ is as in (\ref{eq:lambda}) with $\tau_\theta(x)=\nabla s_\theta(x)$.
\hfill$\Box$

\vspace{0.1in}
\noindent
Hence, if we are using the Stein class for the method of moments, the upshot is that we can ``choose" the $w$ function in (\ref{wscore1}). If we view this $w$ as being equivalent to a transformation then we are effectively ``choosing" $s_\theta(y)$
to minimize (\ref{eq:dhat}) with the transformed $y_{1:n}$ sample; i.e. minimize
\begin{equation}\label{eq:scorey}
\sum_{i=1}^n \left\{s^2_\theta(y_i) +2\,s'_\theta(y_i)\right\}.
\end{equation}
We can achieve a good estimator with a large sample size due to the variance of the terms disappearing. 
If we get to choose the form of $s_\theta(y)$ through the choice of $w$ in (\ref{eq:lambda}) then to reduce the variance we would appeal to $s_\theta(y)$ being constant, or at least the term where $y$ and $\theta$ connect be a constant. For this, the score function could be linear with the coefficient for $\theta$ being a constant, which arises from a normal model. Hence, we want the $w$ to be coherent with a transform aimed at getting the $Y=g(X)$ sample close to normal. For this we will use the Box-Cox class, so for some $\lambda$, $y=(x^\lambda-1)/\lambda$. The corresponding $w$ is given by $w(x)=x^{2(1-\lambda)}$.  

We select the value of $\lambda$ by computing the 
Andersen-Darling statistic for each possible value of $\lambda$ and selecting the value which minimizes the statistics. See \cite{Anderson52}. For the details, for each $\lambda$ we transform
$Z_i=(x_i^{\lambda}-1)/\lambda$ for $i=1,\ldots,n$, and then standardize the data to
$Z_i\to (Z_i-\bar{Z})/S_Z$, where $S_Z$ is the sample standard deviation, and then compute
$$A(\lambda)=-\sum_{i=1}^n (2i-1)\{\log\Phi(Z_{(i)})+\log(1-Z_{(n-i+1)}\}$$
where $(Z_{(i)})$ are the ordered values. We choose the $\lambda$ minimizing $A(\lambda)$.

To elaborate on the point of minimizing variance and the connection with a normal sample, suppose $f(x\mid\theta)$ is a member of the one parameter exponential family. Then following the transformation we would have $s_\theta(y)=\theta\,v(y)+a(y)$ for functions $v$ and $a$, and for which we get to choose $v$. To minimize the variance associated with minimizing (\ref{eq:scorey}) we would like the function $v$ to not change much as the $(y_i)$ change; 
i.e. we want $v$ to be a constant, which would be a normal model. So we want to transform to a normal data. 



\section{GMM for multiparameter exponential family}

We have shown how to derive an estimator based on the Stein class by choices of $w$ and $\tau_\theta$. An alternative way of dealing with the $w$ would be to use a GMM estimator based on multiple choices of $w$. Therefore, here we describe how GMM works for the Stein class with $\tau_\theta=\nabla s_\theta$.
While we focus on exponential family, provided the score function is available, the following will all be able to be implemented though perhaps may depend on numerical methods.
A member of the exponential family is defined by
$f_\theta(x)\propto \exp\{\theta^\top \tilde{v}(x)+\tilde{c}(x)\}$
and therefore $
s_\theta(x)=\partial \log f_\theta(x)/\partial x=\theta^\top v(x)+c(x)=(\theta^\top, 1) s(x)$
with $\tilde{v}'(x)=v(x)$, $\tilde{c}'(x)=c(x)$ and $s(x)^\top=(v(x)^\top,c(x))$.
Hence $\mbox{dim}(\theta)=\mbox{dim}(v(x))$ as well as that of $g_\theta(x)=w(x)\partial s_\theta (x)/\partial
\theta=w(x) v(x)$, and so we obtain a vector function $\lambda$ as in~\eqref{eq:lambda}:
\begin{align*}
\lambda(\theta,x)&= w(x) v(x) s_\theta(x)+ (w(x) v(x))' \\
& = w(x) v(x) \theta^\top v(x)+w(x) v(x) c(x)+ (w(x) v(x))'\\
&= w(x) v(x) v(x)^\top \theta+w(x) c(x) v(x) + (w(x) v(x))' \\
& = A(x) - B(x) \theta,
\end{align*}
where $A(x)=(w(x) v(x))' + w(x) c(x) v(x)$ and $B(x)= - w(x) v(x) v(x)^\top,$ with the multiplication performed component-wise since all but the vector $v(x)$ are scalar functions.
Since $E(\lambda(\theta,X))=0$ we see that $E\{A(X)\}-E\{B(X)\} \theta=0$.
If we are to consider $m$ weight functions, we have $m$ equations of the form
$
A_k-B_k \theta=0,\quad k=1,2,\dots, m,
$
where $A_k=E\{(v w_k)'+w_k c v\}$ and $B_k=E\{v^\top v w_k\}$ with Monte Carlo estimators
$$
\bar{A}_k=\frac{1}{n}\sum_{i=1}^n \{(v(x_i) w_k(x_i))'+w_k(x_i) c(x_i) v(x_i) \}\quad\mbox{and}\quad
\bar{B}_k=\frac{-1}{n}\sum_{i=1}^n v(x_i) v^\top(x_i) w_k(x_i).
$$
Hence, we can use the GMM method (\cite{Andrews}) to estimate $\theta$ by minimizing	
\[
{\cal L}(\theta) = \sum_{k=1}^m \sum_{j=1}^m
\operatorname{tr}\! \left(\bar{A}_k - \bar{B}_k \theta\right)^{\top} W_{jk}
\left(\bar{A}_j - \bar{B}_j \theta\right)=(\mathcal{B}\theta - a)^{\top}W(\mathcal{B}\theta - a)
\]
for some weight matrix \( W \in \mathbb{R}^{pm\times pm}\) with blocks \( W_{jk} \) and
$
\mathcal{B}^\top =
\begin{bmatrix}
\bar{B}_1 , \ldots , \bar{B}_m
\end{bmatrix}
\in \mathbb{R}^{p\times mp},
\quad 
a = \operatorname{vec}(\bar{A}) \in \mathbb{R}^{pm}.
$
Here $W$ is assumed known or estimated from the data.
If $W$ is known, then we can obtain the minimizer of the quadratic function
at
\begin{equation}
\widehat\theta = (\mathcal{B}^{T}W\mathcal{B})^{-1} \mathcal{B}^{T}W a, 
\label{eq:gmm_solution}
\end{equation} 
for invertible $\mathcal{B}^{T}W\mathcal{B}$. 
Otherwise we can use the pseudo-inverse or include a small regularizer, for some
$\widehat\theta = (\mathcal{B}^{T}W\mathcal{B}+\lambda I)^{-1} \mathcal{B}^{T}W a, \quad \lambda > 0.$
Such a solution always exists as $\mathcal{B}^{T}W\mathcal{B}$ is always positive semi-definite if the weights functions are linearly independent and $W$ is positive definite the solution is unique.

In the GMM literature, the optimal choice of $W$ is given by the inverse of the covariance matrix of the moments, i.e.
$W^* = \left(\operatorname{Cov}(\mathcal{B}\theta - a)\right)^{-1}.$
In other words, the optimal weight matrix is the optimal linear combination of all these consistent estimators such that minimal variance $F(\theta)$ is achieved, namely, the chosen $W$ gives more importance to those moments with smaller variance.

As this covariance matrix is unknown and needs to be estimated from the data as well, a valid choice of such an estimator is obtained adopting a two step GMM method to estimate $\theta$, where in the first step we use $W=I_m$ and in the second step we use the estimated covariance matrix of the moments to estimate $W$.
We will not concern ourselves with these details as we only want to illustrate that it is possible to utilize the GMM framework for the case of many weighted score estimators. Since for the exponential family case the weighted score estimator can be expressed as a moment equation for each $j=1,\dots,m$, we can use many such equations with different weight functions to construct a GMM estimator that combines all these equations to produce an overall estimate of $\theta$.

In particular, we can use the sample covariance matrix of the moments to estimate $W$ in a two step procedure: Start with an initial choice of $W$ say identity, then estimate $\theta^{(1)} = (\bar{\mathcal{B}}^{T}W\bar{\mathcal{B}})^{-1} \bar{\mathcal{B}}^{T}W \bar{a}$ by using the Monte Carlo counterparts for $\bar{\mathcal{B}}$ and $\bar{a}$. After obtaining $\theta^{(1)}$, we can	 estimate 
$\widehat{W} = \left((\mathcal{B}\theta^{(1)} - a)(\mathcal{B}\theta^{(1)} - a)^{\top}\right)^{-1}.$
Then use this $\widehat{W}$ to derive the optimal $\theta^{(2)}$ as	
$\theta^{(2)} = (\bar{\mathcal{B}}^{\top}\widehat{W}\bar{\mathcal{B}})^{-1} \bar{\mathcal{B}}^{\top}\widehat{W} \bar{a}.$
This is referred to in the literature as the two step GMM estimator. As $n \to \infty$, $\theta^{(2)}$ converges to the optimal GMM estimator. Iterative convergence procedures can also be used to refine the estimate of $\theta$ further, but in practice as well as in our example, the two step GMM estimator is often sufficient.
As constructed above, this estimator can be seen as a weighted average of many score functions for various weights. In fact, this GMM estimator can be computed in closed form by solving a linear system, which is also computationally efficient.
In a simulation study, one can compare the performance of this GMM estimator with that of a list single weights function score estimator to see if there is any improvement in terms of bias, variance, or mean squared error.

\section{Illustration: Two parameter Gamma  model}

In this section we compare the GMM approach with the optimal choice of weight function from Box-Cox and Andersen-Darling. We demonstrate with the gamma distribution, which is a member of the two parameter exponential family. 
We chose this distribution because the boundary conditions are satisfied for a variety of weight functions and the score function is straightforward to compute.


\begin{figure}[htbp]
    \centering
    \includegraphics[width=15cm,height=8cm]{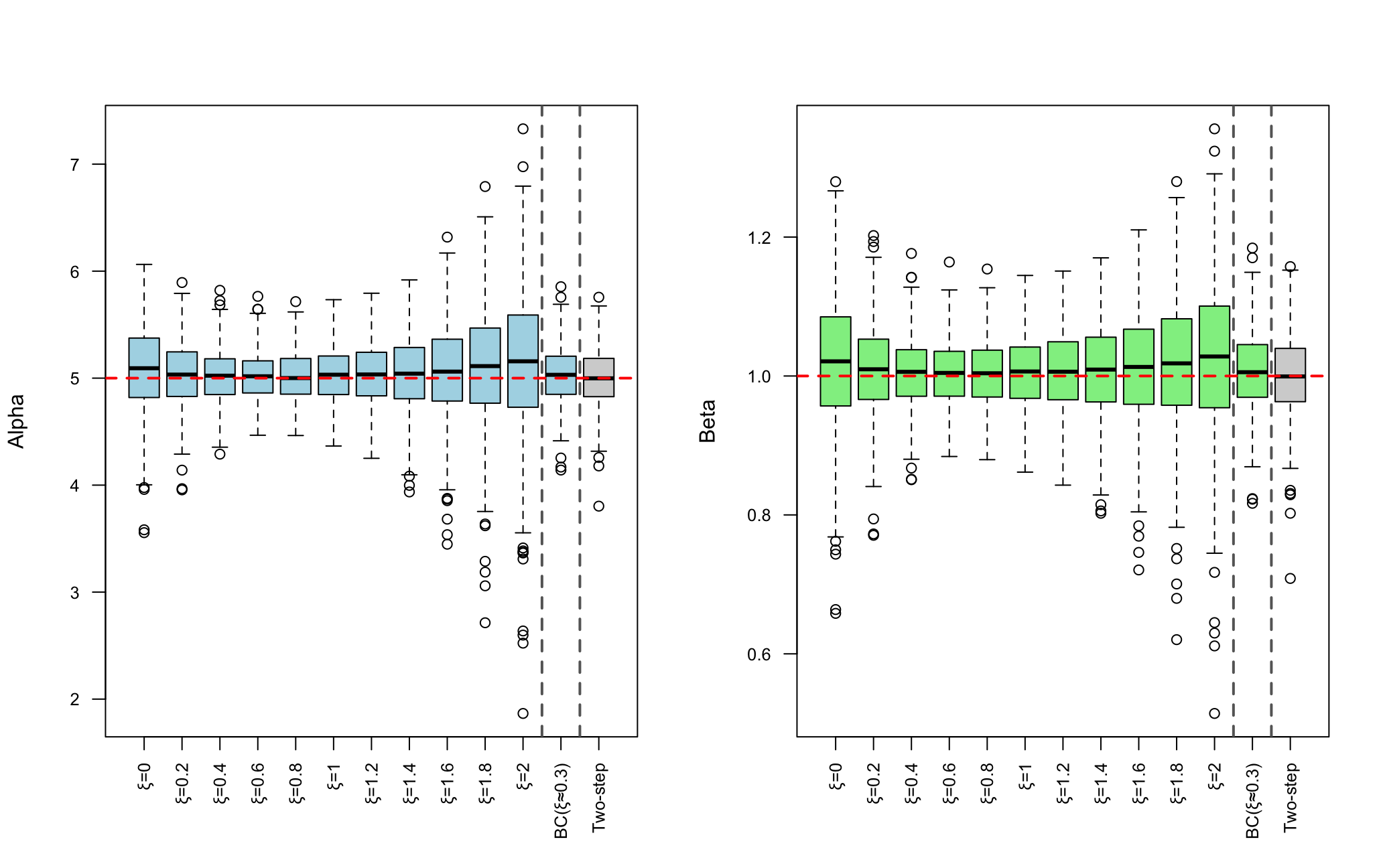}
    \caption{``Good" weight simulation results comparing generalized score matching estimators with the two-step GMM estimator and Box-Cox based estimator. The plots show (left) $\alpha$ estimates as depending on the $(\xi_j)$, (right) $\beta$ estimates depending on the $(\xi_j)$. Red dashed lines indicate true parameter values, and the rightmost boxes in each plot represent the Box-Cox estimators and two-step GMM estimators. }
    \label{fig:simulation1}
\end{figure}

\begin{figure}[htbp]
    \centering
    \includegraphics[width=15cm,height=8cm]{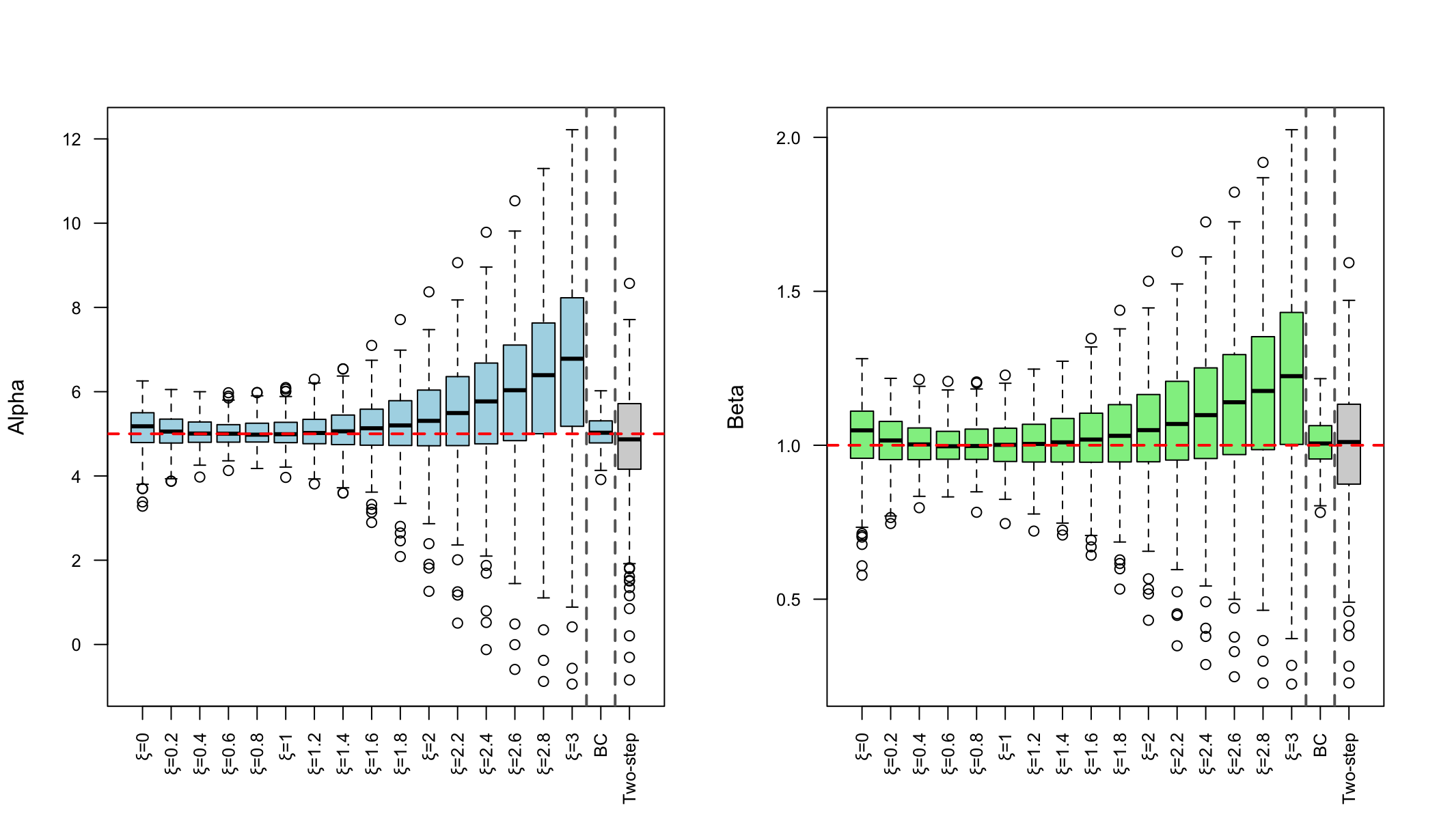}
    \caption{``Bad" weights simulation results comparing generalized score matching estimators with the two-step GMM estimator and Box-Cox based estimator. The plots show (left) $\alpha$ estimates as depending on the $(\xi_j)$, (right) $\beta$ estimates depending on the $(\xi_j)$. Red dashed lines indicate true parameter values, and the rightmost boxes in each plot represent the Box-Cox estimators and two-step GMM estimators. }
    \label{fig:simulation2}
\end{figure}

For the gamma density
$f_\theta(x) = \beta^\alpha\, x^{\alpha-1} e^{-\beta x}/\Gamma(\alpha)$, with $\theta=(\alpha-1,\beta)$,
we have $ 
s_\theta(x)= \theta_1/x - \theta_2$ with $v(x)=(1/x,-1)^\top$ and $c(x)=0$.
The corresponding $A(x)$ and $B(x)$ functions are
$$A(x) = (w(x) v(x))' + w(x) c(x) v(x) = \begin{pmatrix}
-w(x)/x^2 + w'(x)/x \\
-w'(x)
\end{pmatrix},$$
$$B(x) = - w(x) v(x) v(x)^\top = -w(x) \begin{pmatrix}
1/x^2 & -1/x \\
-1/x & 1
\end{pmatrix}.
$$
The Monte Carlo estimates for the entries of $A(x)$ and $B(x)$ are:
$$\bar{A}_{11} = \frac{1}{n}\sum_{i=1}^n w'(x_i)/x_i-\frac{1}{n}\sum_{i=1}^n w(x_i)/x_i^2, \quad
\bar{A}_{12} = \frac{1}{n}\sum_{i=1}^n w'(x_i)$$
\begin{align*} 
\bar{B}_{11} &= -\frac{1}{n}\sum_{i=1}^n w(x_i)/x_i^2, & 
\bar{B}_{12} &= \bar{B}_{21}=\frac{1}{n}\sum_{i=1}^n w(x_i)/x_i, & 
\bar{B}_{22}& = -\frac{1}{n}\sum_{i=1}^n w(x_i) 
\end{align*}
leading to 
\begin{align*}
\begin{pmatrix}
\widehat{\alpha}-1 \\ \\
\widehat{\beta}
\end{pmatrix}
=\begin{pmatrix}
\bar{B}_{11} & \bar{B}_{12} \\ \\
\bar{B}_{21} & \bar{B}_{22}
\end{pmatrix}^{-1}
\begin{pmatrix}
\bar{A}_{11} \\ \\
\bar{A}_{12}
\end{pmatrix}=\begin{pmatrix}
\frac{\bar{B}_{22}\bar{A}_{11} - \bar{B}_{12}\bar{A}_{12}}{\bar{B}_{11}\bar{B}_{22} - \bar{B}_{12}\bar{B}_{21}} \\ \\
\frac{\bar{B}_{11}\bar{A}_{12} - \bar{B}_{21}\bar{A}_{11}}{\bar{B}_{11}\bar{B}_{22} - \bar{B}_{12}\bar{B}_{21}}
\end{pmatrix}.
\end{align*} 


Fig.~\ref{fig:simulation1} shows the results from our simulation study comparing different GMM estimation methods with the original ``good" weight functions and also with the Box-Cox motivated choice of optimal weight function. The weight functions chosen are 
$w_j(x)=x^{\xi_j}$ for $\xi=0,0.3,0.4,0.5,0.8,1.0,1.2,1.5,1.8,2.0$.
Recall the weight function from the Box-Cox procedure would be of the form $w(x)=x^{2(1-\lambda)}$.

In particular, we run 1000 simulations for sample size of $n=500$ from a gamma(5,1) distribution. 
The case of $\xi=0$ corresponds to the standard score matching estimator without weighting. 
Each boxplot represents the distribution of parameter estimators ($\alpha$ and $\beta$) across the 1000 simulations for each $\xi$ value, as well as for the two-step GMM estimator that combines all $\xi$ values and the Box-Cox based estimator with the data driven $\lambda$s based on Box-Cox and the Andersen-Darling statistics.

As shown in Fig.~\ref{fig:simulation1}, the Box-Cox estimator and the two-step GMM estimator (rightmost box in coral color) perform about the same. Given the simplicity of the Box-Cox procedure and the general applicability we would recommend this over GMM, particularly for non exponential family models where the GMM approach would be particularly complicated to implement.

On the other hand Fig.~\ref{fig:simulation2} shows the same results when this time some ``bad" weights are included for the GMM. In this study we extended the powers for the weight functions to include powers between 2 and 3. Despite this apparently harmless extension it is seen that the GMM estimator is very sensitive to the choice of weights; the GMM estimator performing quite poorly. While we label the Fig.~\ref{fig:simulation2} as ``bad" weights the point is that this could not be known upfront. 

\section{Weibull model and real data illustration}

In this section we apply our method to a non exponential family model, the Weibull distribution. We compare the performance of the Box-Cox based estimator  with a range of weight functions as well the maximum likelihood estimator.
The score function and the corresponding weighted score matching objective function for the Weibull distribution is derived in the following. Note tha the optimal solution here needs to be obtained through numerical optimization, as the score function is now not linear in the parameters.






A random variable $X$ follows a Weibull distribution with shape $k > 0$ and scale $\kappa > 0$ if the probability density function is given by
\[ f(x; \theta) = \frac{k}{\kappa} \left( \frac{x}{\kappa} \right)^{k-1} \exp\left( -\left( \frac{x}{\kappa} \right)^k \right) \]
and $\theta = (k, \kappa)$ is the parameter vector.
The log-density is given by:
$ \ln f(x; \theta) = \ln k - k \ln \kappa + (k-1) \ln x - \left( x/\kappa \right)^k $
and the score function $s_\theta(x)$ is defined as the derivative with respect to $x$; i.e.
$ s_\theta(x) = \partial \ln f/\partial x = (k-1)/x- k x^{k-1}/\kappa^k $
and the derivative of the score function is given by
$ s_\theta'(x) = -(k-1)/x^2 - k(k-1)x^{k-2}/\kappa^k. $
To estimate $\theta = (k, \kappa)$, we minimize the weighted Fisher divergence $J_\lambda(\theta)$ using the weight function $w(x) = x^{2-2\lambda}$. The objective function is:
\[ J_\lambda(\theta) = \frac{1}{n} \sum_{i=1}^n \left[ w(X_i) \left( \frac{1}{2} s_\theta(X_i,\theta)^2 + s_\theta'(X_i, \theta) \right) + w'(X_i) s_\theta(X_i, \theta) \right] \]
where the weight function and its derivative are
$w(x) = x^{2-2\lambda}$ and $w'(x) = (2-2\lambda)x^{1-2\lambda}$.
Substituting the Weibull components into the objective function, we analyze the expression term by term.

\begin{description}

\item 1: $\frac{1}{2} w(x) s_\theta(x)^2$
\[ \frac{1}{2} x^{2-2\lambda} \left( \frac{k-1}{x} - \frac{kx^{k-1}}{\kappa^k} \right)^2 = \frac{(k-1)^2}{2} x^{-2\lambda} - \frac{k(k-1)}{\kappa^k} x^{k-2\lambda} + \frac{k^2}{2\kappa^{2k}} x^{2k-2\lambda} \]

 \item 2: $w(x) s_\theta'(x)$
\[ x^{2-2\lambda} \left( -\frac{k-1}{x^2} - \frac{k(k-1)x^{k-2}}{\kappa^k} \right) = -(k-1) x^{-2\lambda} - \frac{k(k-1)}{\kappa^k} x^{k-2\lambda} \]

\item 3: $w'(x) s_\theta(x)$
\[ (2-2\lambda)x^{1-2\lambda} \left( \frac{k-1}{x} - \frac{kx^{k-1}}{\kappa^k} \right) = 2(1-\lambda)(k-1)x^{-2\lambda} - \frac{2(1-\lambda)k}{\kappa^k} x^{k-2\lambda} \]

\end{description}

\noindent
Grouping the terms by the powers of $X_i$, the final objective function simplifies to:
\[ J_{\lambda}(k, \kappa) = \frac{1}{n} \sum_{i=1}^n \left[ C_1 X_i^{-2\lambda} + C_2 X_i^{k-2\lambda} + C_3 X_i^{2k-2\lambda} \right] \]
where the coefficients are defined as:
 $   C_1 = (k-1)(k + 1 - 4\lambda)/2$, 
    $C_2 = -2k(k-\lambda)/\kappa^k $  and
    $C_3 = k^2/(2\kappa^{2k})$.
For the score matching estimator to be consistent, the boundary condition $h(x)f(x)\psi(x) \to 0$ must hold as $x \to 0$. For the Weibull distribution this requires
$ k > 2\lambda$.
For fixed $\lambda$ and observations $x_1,\dots,x_n$, define
\[
A_0=\frac1n\sum_{i=1}^n x_i^{-2\lambda},\qquad
A_1(k)=\frac1n\sum_{i=1}^n x_i^{k-2\lambda},\qquad
A_2(k)=\frac1n\sum_{i=1}^n x_i^{2k-2\lambda}.
\]
Then
\[
J_\lambda(k,\kappa)=C_1A_0-\frac{2k(k-\lambda)}{\kappa^k}A_1(k)+\frac{k^2}{2\kappa^{2k}}A_2(k).
\]
The $k$-derivative summaries are
\[
B_1(k)=\frac1n\sum_{i=1}^n x_i^{k-2\lambda}\ln x_i,\qquad
B_2(k)=\frac1n\sum_{i=1}^n x_i^{2k-2\lambda}\ln x_i,
\]
so that
$
A_1'(k)=B_1(k),\qquad A_2'(k)=2B_2(k).
$
The gradient w.r.t. $\kappa$ is given by
\[
\frac{\partial J_\lambda}{\partial \kappa}
=2k^2(k-\lambda)A_1(k)\kappa^{-k-1}-k^3A_2(k)\kappa^{-2k-1}.
\]
Setting this to zero gives
\[
\kappa^k=\frac{kA_2(k)}{2(k-\lambda)A_1(k)}
\quad\Rightarrow\quad
\kappa^*(k)=\left[\frac{kA_2(k)}{2(k-\lambda)A_1(k)}\right]^{1/k},
\]
(which requires $k>\lambda$ for positivity).
The gradient w.r.t. $k$ is given by
\[
\begin{array}{ll}
\frac{\partial J_\lambda}{\partial k}
& =(k-2\lambda)A_0
-2\kappa^{-k}\Big((2k-\lambda)-k(k-\lambda)\ln\kappa\Big)A_1(k) \\
& -2k(k-\lambda)\kappa^{-k}B_1(k)
+\kappa^{-2k}\big(k-k^2\ln\kappa\big)A_2(k)
+k^2\kappa^{-2k}B_2(k).
\end{array}
\]
Hence, minimization can be done by profiling
$
k^*=\arg\min_{k>\lambda} J_\lambda\big(k,\kappa^*(k)\big)$,
$\kappa^*=\kappa^*(k^*).
$
In practice, we solve the one-dimensional equation
$
(\partial /\partial k)\,J_\lambda\big(k,\kappa^*(k)\big)=0
$
numerically (e.g., Newton or line search), then back-substitute for $\kappa^*$.

\begin{figure}[htbp]
    \centering
    \begin{minipage}{0.7\textwidth}
        \centering
        \includegraphics[width=\linewidth]{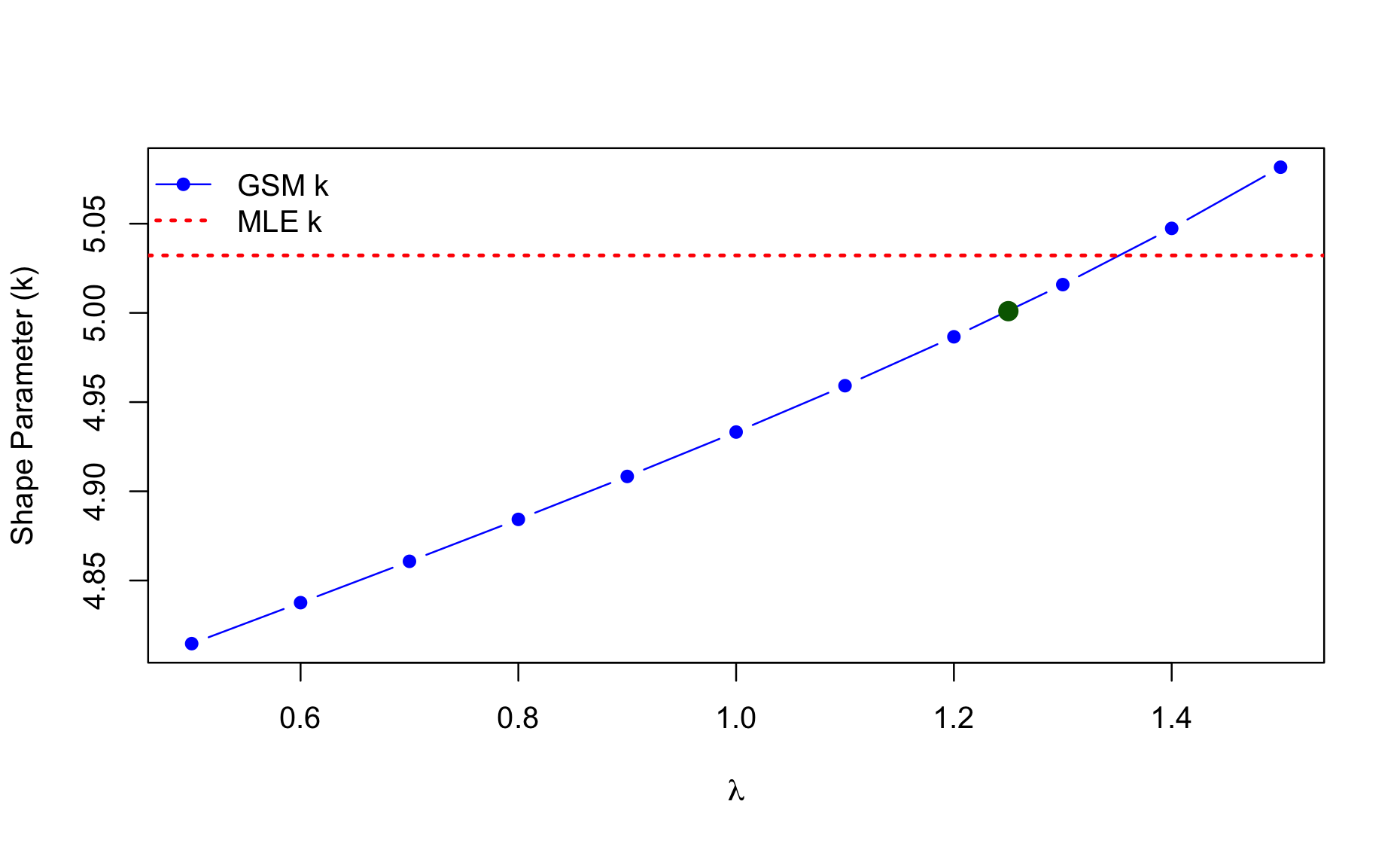}
    \end{minipage}\hfill
    \begin{minipage}{0.7\textwidth}
        \centering
        \includegraphics[width=\linewidth]{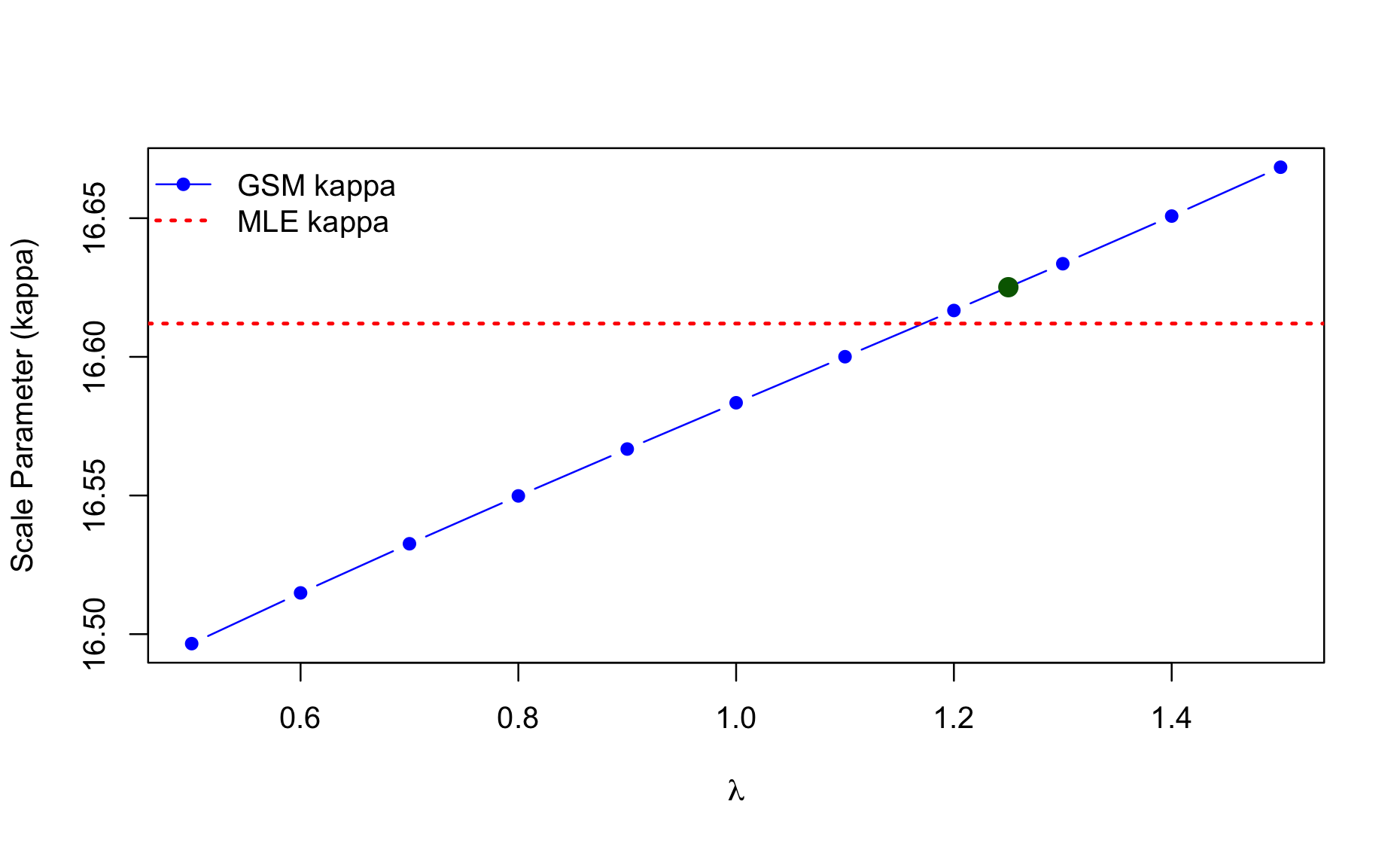}
    \end{minipage}
    \caption{MLE for the Weibull parameters (red line) alongside generalized score matching estimators for varying $\lambda$ (blue dots). The green dot is the Box-Cox based estimator which selects a value for $\lambda$}
    \label{fig:weibull_estimators}
\end{figure}

We apply the above method to a real dataset, the ``TreesDBH" data set which comes from the R package ``WeibullFit" and consists of 50,607 diameters of Brazilian eucalyptus trees measured at 4.5 feet off the ground. This is a standard way for measuring trees.
We wrote our own MLE code which is the same as that from the R package ``Weibullness".  The analysis of estimators from various settings in presented in Fig.~\ref{fig:weibull_estimators}. The upper plot is concerned with the parameter $k$ and the lower plot is for $\kappa$. The red line in each case is the MLE. 
Each blue dot represents the estimator based on the given value of $\lambda$, i.e. taking the weight function to be $w(x)=x^{2(1-\lambda)}$. The green dot is the estimator based on the Box-Cox transformation which selects a value of $\lambda$. As can be seen, the Box-Cox based estimator is very close to the MLE.
The sample size is quite large, $n=50,607$, which explains the small $y$-axes scales. Indeed, for such a large $n$, it is safe to assume that $\mbox{Var}(\widehat\kappa)$ and $\mbox{Var}(\widehat{k})$ will be approximately normal with means the true $\kappa$ and $k$, respectively, and variance the reciprocal of $n$ times the respective Fisher information values evaluated at the MLE values. So we estimate 2 times the standard deviations of the MLEs as 0.034 and 0.030 for $\widehat{k}$ and $\widehat\kappa$, respectively. The Box-Cox estimator is within this interval while many of the other estimators are not.

\section{Theoretical results}

First we look at the convergence of the Box-Cox estimated value of $\lambda$.
From the observed data $X_{1:n}$ we construct 
$$Y_i^\la=(X_i^\lambda-\bar{X}^\la)/S^\la,\quad i=1,\ldots,n,$$
for $\lambda\in U$, and $U$ is a compact set of values, where $\bar{X}^\la$ and $S^\la$ are the sample mean and the sample variance of the $(X_i^\lambda)$, respectively. Note that we would get the same values of $(Y_i^\la)$ if we had used the full definition of the Box-Cox transform, i.e. $(X^\lambda-1)/\lambda$. The $\lambda_n$ minimizes
$d(F_{n\lambda},\Phi)$ where
$$d(F,\Phi)=\int\frac{(F(x)-\Phi(x))^2}{\Phi(x)(1-\Phi(x))}\,\phi(x)\,dx,$$
where $\phi$ and $\Phi$ are the density and distribution for the standard normal.
The $F_{n\lambda}$ is the empirical distribution function of the $(Y_i^{\la})$ and so
$$F_{n\lambda}(y)=n^{-1}\sum_{i=1}^n 1\left(X_i\leq (yS^\la+\bar{X}^\la)^{1/\lambda}\right).$$
Hence,
$F_{n\lambda}(y)=F_{nX}\big((yS^\la+\bar{X}^\la)^{1/\lambda}\big),$
where $F_{nX}$ is the empirical distribution function of the $(X_i)$.
We can use \cite{Andrews1987} to show that
\begin{equation}\label{limit}
\sup_{\lambda\in U}\left|\bar{X}^\la-\mu^\la\right|\to 0\quad\mbox{and}\quad \sup_{\lambda\in U}\left|S^\la-\sigma^\la\right|\to 0,
\end{equation}
almost surely (or in probability, see condition 2. below), where $\mu^\la=\int x^\lambda\,dF(x)$ and $\sigma^\la=\sqrt{\int (x^{\lambda})^2\,dF(x)-(\mu^\la)^2}$. Here $U$ is of the form $[\lambda_1,\lambda_2]$.
The assumptions from \cite{Andrews1987} for (\ref{limit}) to hold are that
\begin{description}

\item 1. $\lambda\in U$ with $U$ compact.

\item 2. It is that $$\left|n^{-1}\sum_{i=1}^n X_i^{\tau\lambda}-E(X^{\tau\lambda})\right|\to 0\quad\mbox{a.s.}\quad (\mbox{in probability})$$
for all $\lambda\in U$ for both $\tau=1$ and $\tau=2$.

\item 3. It is that
$$\lim_{\rho\to 0}
\left| E\{X^{\tau(\lambda+\rho)}\}-E\{X^{\tau\lambda}\}\right|=0
$$
for all $\lambda\in U$ and for both $\tau=1$ and $\tau=2$.

\end{description}

\noindent
These assumptions depend solely on the distribution of the $X$, i.e. $F$. 

\noindent
\begin{theorem}
With conditions 1., 2. and 3.,
\begin{equation}\label{limdist}
\sup_{y,\lambda}\left|F_{n,\lambda}(y)-F_\lambda(y)\right|\to 0\quad\mbox{a.s.}
\end{equation}
where 
$F_\lambda(y)=F\big((y\sigma^\la+\mu^\la)^{1/\lambda}\big).$
Further,
$\sup_{\lambda\in U} \left|d(F_{n,\lambda},\Phi)-d(F_{\lambda},\Phi)\right|\to 0$ a.s.
which ensures $\lambda_n\to \lambda_0$ a.s., where $\lambda_0$ minimizes $d(F_\lambda,\Phi)$.  
\end{theorem}

\noindent
{\sl Proof}.
The proof to this starts with defining
$$F_{n\lambda}^*(y)=n^{-1}\sum_{i=1}^n 1\left(X_i\leq (y\sigma^\la+\mu^\la)^{1/\lambda}\right)$$
which is 
$F_{nX}\left((y\sigma^\la+\mu^\la)^{1/\lambda}\right).$
So from the usual convergence of empirical distribution functions it is that 
$\sup_{y,\lambda}|F_{n\lambda}^*(y)-F_{\lambda}(y)|\to 0\quad\mbox{a.s.}$ (see Lemma 3 just after this proof).
Also, using (\ref{limit}), we have
$\sup_{y,\lambda}|F_{n\lambda}(y)-F_{n\lambda}^*(y)|\to 0\quad\mbox{a.s.}$
yielding
$\sup_{y,\lambda}|F_{n\lambda}(y)-F_\lambda(y)|\to 0\quad \mbox{a.s.}$
which in turn implies (\ref{limdist}) from the continuity of $d(F,\Phi)$.
\hfill$\Box$

\vspace{0.1in}
\noindent
The following lemma covers the relevant part of the proof to theorem 1. 

\begin{lemma}
If $F_n(y)=n^{-1}\sum_{i=1}^n 1(X_i\leq y\psi_n)$ and $F^*_n(y)=n^{-1}\sum_{i=1}^n 1(X_i\leq y\psi)$ where
$\psi_n\to \psi$, with $\psi_n>0$, and the $(X_i)$ are i.i.d. $F$ and positive, then
$\sup_y|F_n(y)-F_n^*(y)|\to 0$.
\end{lemma}

\noindent
{\sl Proof}.
Without loss of generality, assume that $\psi_n>\psi$ for all large $n$, then
$$1(X_i\leq y\psi_n)-1(X_i\leq y\psi)=1(y\psi\leq X_i\leq y\psi_n)$$
and so it is easy to see that we only need to consider $X_{(1)}/\psi_n<y<X_{(n)}/\psi$, where $(X_{(i)})$ are the ordered $(X_i)$. 
Hence, for the result, we just need $X_{(n)}(\psi_n-\psi)\to 0$ as $n\to\infty$. Since $\psi_n-\psi$ will behave as $n^{-1/2}$ and
$X_{(n)}$ as $\log n$, the proof is complete.
\hfill$\Box$

\vspace{0.2in}
\noindent
We estimate $\theta$ with a sample of size $n$ by minimizing
$$L_n(\theta)=n^{-1}\sum_{i=1}^n \left\{w_n(X_i)\,s_\theta^2(X_i)+2(s_\theta(X_i)w_n(X_i))'\right\},$$
where $w_n(x)=x^{2(1-\lambda_n)}$. 

\begin{theorem}
It is that 
$\sup_{\theta\in\Theta}|L_n(\theta)-L_{n,0}(\theta)|\to 0\quad\mbox{in probability},$
where 
$$L_{n,0}(\theta)=n^{-1}\sum_{i=1}^n \left\{w_0(X_i)\,s_\theta^2(X_i)+2(s_\theta(X_i)w_0(X_i))'\right\},$$
and $w_0(x)=x^{2(1-\lambda_0)}$.
\end{theorem}

\noindent
{\sl Proof}.
For $\theta\in\Theta$, define
$$L_n(\theta)-L_{n,0}(\theta)=n^{-1}\sum_{i=1}^n \left\{\delta_n(X_i)s_\theta^2(X_i)+2s'_\theta(X_i)\delta_n(X_i)+2s_\theta(X_i)\delta'_n(X_i)\right\},$$
where $\delta_n(x)=w_n(x)-w_0(x)$.
To show that $L_n$ converges uniformly to $L_0$ we will deal with each of the three terms and use the H\"older inequality.
So let $\epsilon_n(x)$ represent $\delta_n(x)$ and $\delta'_n(x)$, while $g_\theta(x)$ represents $s^2_\theta(x)$ and $s'_\theta(x)$.
If for some $\psi>0$ it is that
\begin{equation}\label{ass1}
\sup_{\theta\in\Theta} n^{-1}\sum_{i=1}^n g_\theta(x_i)^{1+\psi}<M
\end{equation}
in probability for some $M<\infty$ and
\begin{equation}\label{ass2}
n^{-1}\sum_{i=1}^n \epsilon_n(x_i)^{1+1/\psi}\to 0\quad\mbox{almost surely}
\end{equation}
then $L_n$ converges uniformly to $L_0$.
The H\"older inequality is applied to
$\sup_{\theta\in\Theta}n^{-1}\sum_{i=1}^n \epsilon_n(X_i)\,g_\theta(X_i),$
i.e.
$$n^{-1}\sum_{i=1}^n \epsilon_n(X_i)g_\theta(X_i)\leq\left(n^{-1}\sum_{i=1}^n \epsilon_n(X_i)^{1+1/\psi}\right)^{1/(1+1/\psi)}\left(n^{-1}\sum_{i=1}^n g_\theta(X_i)^{1+\psi}\right)^{1/(1+\psi)}$$
with $1/(1+1/\psi)+1/(1+\psi)=1$. The result follows with (\ref{ass1}) and (\ref{ass2}).
\hfill$\Box$

\noindent
The uniform convergence of $L_n(\theta)-L_{n,0}(\theta)$ to 0 and the uniform convergence of $L_{n,0}(\theta)$ to $L_0(\theta)$, 
where 
$$L_0(\theta)=\int w_0(x)\,\{s_\theta(x)-s(x)\}^2\,f(x)\,dx,$$
implies $\theta_n$, the minimizer of $L_n(\theta)$ converges to $\theta_0$, the minimizer of 
$L_0(\theta)$.
Hence $\theta_0$ is the true value of $\theta$.

We can consider the asymptotic normality of $\theta_n$, the maximizer of $L_n(\theta)$. To this end we start with the derivative of $L_n(\theta)$ with respect to $\theta$, i.e. $L_n'(\theta)$ and so $L_n'(\theta_n)=0$.  
Then using a Taylor expansion about $\theta_0$, the true parameter value,
$$0=L_n'(\theta_0)+(\theta_n-\theta_0)L_n''(\theta)+\half (\theta_n-\theta_0)^2\,L_n'''(\widetilde{\theta}_n)$$
for some $\widetilde{\theta}_n$ between $\theta_n$ and $\theta_0$.
Hence,
$$\theta_n-\theta_0=\frac{-L_n'(\theta_0)}{L_n''(\theta_0)+\half(\theta_n-\theta_0)L_n'''(\widetilde\theta_n)}.$$
Using Chapter 5 from the book Asymptotic Statistics (\cite{Vaart1998}), we, as does van der Vaart,  assume that $(\theta_n-\theta_0)L_n'''(\widetilde\theta_n)$ converges to 0, based on $\theta_n\to\theta_0$ and the assumed boundedness of $|L_n'''(\widetilde\theta_n)|$.  Most results for the asymptotic normality for method of moment estimators assume that
$$L_n'(\theta_0)=n^{-1}\sum_{i=1}^n \Psi_{\theta_0}(X_i)$$
and then there is an easy application of a central limit theorem for
$\sqrt{n}\,L_n'(\theta_0)$. The denominator can be dealt with by a law of large numbers and so will converge to a constant and the asymptotic normality of $\theta_n$ follows. However, our $L_n'(\theta_0)$ is slightly more complicated and involves three terms which we write as
$$L_n'(\theta_0)=n^{-1}\sum_{i=1}^n \gamma(w_n(X_i),\,s_{\theta_0}(X_i))$$
for the linear in $w_n$ function $\gamma$; i.e.,
$\gamma(w,s)=(\partial/\partial\,\theta)\,(ws_\theta^2+2(w\,s_\theta)')$.

The obvious way to proceed is to write
$$L_n'(\theta_0)=n^{-1}\sum_{i=1}^n \gamma\big(w_n(X_i)-w_0(X_i), s_{\theta_0}(X_i)\big)+L_{n,0}'(\theta_0).$$
The second term on the right multiplied by $\sqrt{n}$ can be shown to be asymptotically normal, following \cite{Vaart1998}, while the first term on the right multiplied by $\sqrt{n}$ will be shown to converge to 0.

First, and easiest, $\sqrt{n}\,L'_{n,0}(\theta_0)$ is asymptotically normal with mean 0 and variance
$$V=\mbox{Var}\bigg(w_0(X)\,s_{\theta_0}(X)+2(w_0(X)\,s_{\theta_0}(X))'\bigg)$$
with $X$ coming from the true density function.

Second, assuming that $|\lambda_n-\lambda_0|=O_p(1/\sqrt{n})$, which is standard for the convergence of one dimensional parameters, we can show that
$\sqrt{n}(w_n(x)-w_0(x))\to w_0(x)\,\log x$.
Let us now write
$2(1-\lambda_n)=\alpha_n$ and $2(1-\lambda_0)=\alpha_0$ and assume we can write $\alpha_n=\alpha_0+c/\sqrt{n}$ for some finite $c$. That it is random and approximately normal is not relevant to what follows.
We are now looking for the result that if
$$g_n(x)=x^{\alpha_0}\left\{\sqrt{n}\left(\exp\{\log (x)/\sqrt{n}\}-1 \right)-\log x\right\}$$
then 
$$n^{-1}\sum_{i=1}^n g_n(X_i)\to 0.$$
This result is available if there exists a $M_n\to\infty$ for which
$\sup_{x<M_n}g_n(x)\to 0$ and $\sum_{n=1}^\infty P(X>M_n)<\infty$.
It is easy to see that the sup condition will be satisfied for
$M_n=n^\tau$ for $\tau<\alpha_0/2$ if $\alpha_0>0$. We effectively require $n^{\tau\alpha_0}-1/2\to 0$ as $n\to\infty$ and so is automatic if $\alpha_0<0$. We also need such a result for the derivative of $\sqrt{n}(w_n(x)-w_0(x))$ leading to
$\tau<(\alpha_0-1)/2$ if $\alpha_0>1$ whereas $\tau<\alpha_0/2$ if $0<\alpha_0<1$ and $\tau$ has no restriction if $\alpha_0<0$.
For such $\tau$ we require $\sum_{n=1}^\infty P(X>n^\tau)<\infty$ which will hold for all tails satisfying $P(X>x)<x^{-(1+\delta)/\tau}$ for some $\delta>0$.
Hence, under the above conditions, we can show that
$$\lim_{n\to\infty}\left| n^{-1/2}\sum_{i=1}^n \gamma\bigg(w_n(X_i)-w_0(X_i), s_{\theta_0}(X_i)\bigg)\right|\leq  2\int w_0(x)\,|\log x|\,(s_{\theta_0}(x)-s(x))\,f(x)\,dx=0,$$
since $s_{\theta_0}(x)=s(x)$.

Finally, to pin down the asymptotic variance, we have from a law of large numbers that $L_n''(\theta_0)$ converges to the constant $\xi>0$.
So overall $\sqrt{n}(\widehat\theta_n-\theta_0)$ is asymptotically normal with mean $0$ and variance $V/\xi^2$.

A summary of what has just been gone through goes as follows:
If we were looking at the problem of optimizing
$n^{-1}\sum_{i=1}^n \Psi_\theta(X_i)$
then the techniques for showing consistency and asymptotic normality are standard, see \cite{Vaart1998}, for example. However, we  have
$n^{-1}\sum_{i=1}^n \Psi_{n,\theta}(X_i)$
and $\Psi_{n,\theta}(X)\to \Psi_\theta(X)$.
Hence, we write
$$n^{-1}\sum_{i=1}^n \Psi_{n,\theta}(X_i)=n^{-1}\sum_{i=1}^n\left\{\Psi_{n,\theta}(X_i)-\Psi_\theta(X_i)\right\}+n^{-1}\sum_{i=1}^n \Psi_{\theta}(X_i).$$
We go on to show that the first term on the right is negligible and the second term on the right can be dealt with using the standard techniques in the literature. 









\section{Discussion}

We have shown that the score matching and generalized  score matching estimators can be placed within a class of Stein method of moment estimator. 
By placing generalized score matching estimators within the generalized method of moments framework we have been able to work with multiple weight functions and derive an optimal estimator. 

While we have focused on the two parameter gamma distribution, the possible examples are limitless, and would involve applications of GMM with individual estimators being generalized score matching estimators. Such illustrations presented here would add no new insights with the key connect being that the $\tau_\theta$ in (\ref{eq:lambda}) is taken to be $\nabla s_\theta$ where $s_\theta$ is the score function.

For example, if 
$$f(x\mid\theta)=\exp\left( \sum_{j=1}^p \theta_j\,\phi_j(x)\right),$$
where the $(\phi_j)$ are some suitable set of functions, then
$\tau_\theta(x)$ is the $p$-vector of functions $(\phi'_j(x))$.
Hence, $\lambda(x,\theta)$ is a $p$ dimensional function with component functions
$\lambda_j(x,\theta)=s_\theta(x)\,w(x)\,\phi'_j(x)+(w(x)\,\phi'_j(x))'.$
The $p$-dimensional estimator satisfies $\sum_{i=1}^n \lambda_j(x_i,\theta)=0$ for all $j=1,\ldots,p$.
The estimator is given by $\widehat\theta=M^{-1}v$ where $v$ is a $p$-vector and $M$ a $p\times p$ matrix with elements
$$v_j=-\sum_{i=1}^n w(x_i)\,\phi_j(x_i))'\quad\mbox{and}\quad M_{jk}=\sum_{i=1}^n w(x_i)\,\phi_j'(x_i)\,\phi_k'(x_i).$$
From a collection of weight functions $(w_j)$ we can then obtain the GMM estimator.

\begin{description}

\item 1. Stein MM uses
$$\lambda(x,\theta)=\frac{(w(x)\,\tau_\theta(x)f_\theta(x))'}{f_\theta(x)}.$$

\item 2. GMM uses multiple $w_j(x)$ and averages. 

\item 3. Take $\tau_\theta(x)=(\partial/\partial\theta)s_\theta(x)$ so that the estimator coincides with generalized score matching.

\item 4. Treat $w$ as a transform so we recover a score matching estimator.

\item 5. We can choose $s_\theta(y)$ and minimize
$$\sum_{i=1}^n \{s^2_\theta(y_i)+2s'_\theta(y_i)\}.$$

\item 6. The best estimation for a parameter, because it is MLE, occurs when the $(y_i)$ are normal. 

\item 7. Hence, use transform from $x\to y$ to get close to normal, using e.g. Box-Cox.

\end{description}

\noindent
Example: If
$s_\theta(x)=\theta v'(x)+a(x)$
then 
$$s_\theta(y)=\theta v'(g(y))\,g'(y)+g''(y)/g('y)+a(g(y))g'(y).$$
The score for a normal is of the form $c\theta+l(y)$
where $l$ is a linear function in $y$.
To minimize the variance of the problem, replicating a large sample, we want $v'(g(y_i))g'(y_i)$ to lack variability; i.e. to be a constant. This is when $(v(g(y))'$ is constant, i.e. $v(g)$ is linear, which implies normal. Pick $g$ accordingly. 




  \bibliography{score.bib}

@article{Andrews1987,
author={Andrews, D.W.K.},
title={Consistency in Nonlinear Econometric Models: A Generic Uniform Law of Large Numbers},
journal={Econometrica}, 
volume={55}, 
year={1987}, 
pages={1465-1471}
}

@book{Vaart1998,
author={van der Vaart, A.W.},
title={Asymptotic Statistics},
year={1998},
publisher={Cambridge University Press}
}

@incollection{Oates22,
	title ={Minimum kernel discrepancy estimators},
	editor={Hinrichs, A. and Kritzer, P. and Pillichshammer, F.},
	booktitle = {Monte Carlo and Quasi Monte Carlo Methods},
	year ={2022},
	publisher={Springer Verlag},
	author ={Oates, C.}
}

@article{Barp19,
	author={Barp, A. and Briol, F. and Duncan, A.B. and Girolami, M. and Mackey, L.},
	title={Minimum Stein discrepancy estimators},
	year={2019},
	journal={33rd NeurIPS Conference Proceedings}
}

@article{Stein72,
author={Stein, C.}, 
year={1972}, 
title={A bound for the error in the normal approximation to the distribution of a sum of dependent random variables.},
journal={Proceedings of the Sixth Berkeley Symposium on Mathematical Statistics and Probability}, 
volume={2},
pages={583–602}
}

@article{quandt78,
author={Quandt, R.E. and Ramsey, J.B.}, 
year={1978}, 
title={Estimating mixtures of normal distributions and switching regressions}, 
journal={Journal of the American Statistical Association}, 
volume={73}, 
pages={730–752}
}

@article{Ebner,
author={Ebner, B. and Fischer, A. and Gaunt, R.E. and Picker, B. and Swan, Y.},
title={Stein's method of moments},
year={2024},
journal={arXiv:2305.19031v5.}
}

@article{Andrews,
author={Andrews, D.W.K.},
title={Consistent moment selection procedures for
      generalized method of moments estimation},
journal={Econometrica},
year={1999},
volume={67},
pages={543-564}
}

@book{Hall05,
author={Hall, A.R.},
title={Generlaized Method of Moments},
publisher={Oxford University Press},
year={2005}
}

@article{Hansen82,
author={Hansen, P.},
title={Large sample properties of generalized method of moments estimators},
journal={Econometrica},
year={1982},
volume={50},
pages={1029-1054}
}

@article{Atkinson1981,
author={Atkinson, Colin},
title={Rao's Distance Measure}, 
journal={Sankhya: The Indian Journal of Statistics, Series A}, 
year={1981},
volume={43},
pages={345–365}
}

@article{Box64,
	author={Box, G.E.P. and Cox, D.R.},
	title={An analysis of transformations},
	journal={Journal of the Royal Statistical Society, Series B},
	volume={26},
	year={1964},
	pages={211-252}
}

@article{Hyvarinen05,
	author={Hyvarinen, A},
	title={Estimation of non-normalized statistical models by score matching},
	journal={Journal of Machine Learning Research},
	year={2005},
	volume={6},
	pages={695-709}
}

@article{Anderson52,
	author={Anderson, T.W. and Darling, D.A.}, 
	year={1952}, 
	title={Asymptotic theory of certain goodness-of-fit criteria based on stochastic processes}, 
	journal={Annals of Mathematical Statistics}, 
	volume={23}, 
	pages={193–212}
}

@article{Xu25,
	author={Xu, J. and Scealy, J.L. and Wood, A.T.A. and Zou, T.},
	title={Generalized score matching},
	journal={Journal of Multivariate Analysis},
	year={2025},
	volume={210},
	pages={105473}
}

@article{Yu19,
	author={Yu, S. and Drton, M. and Shojaie, A.},
	journal={Journal of Machine Learning Research},
	title={Generalized score matching for non-negative data},
	year={2019},
	volume={20},
	pages={1-70}
}

@article{Scealy23,
	author={Scealy, J.L. and Wood, A.T.A.},
	title={Score matching for compositional distributions},
	journal={Journal of the American Statistical Association},
	year={2023},
	volume={118},
	pages={1811-1823}
}

\end{document}